\documentclass{article}
\usepackage{arxiv} % see https://github.com/kourgeorge/arxiv-style
\usepackage{graphicx} % Required for inserting images
\usepackage{url}
\usepackage{hyperref}
\usepackage{orcidlink}
\usepackage{csquotes}

\title{Datenschutzkonformer LLM-Einsatz: Eine Open-Source-Referenzarchitektur}
\author{Marian Lambert \orcidlink{0000-0001-7347-5734} \\
    XPACE GmbH\\
    \texttt{marian.lambert@xpace.de}
    \And
    Thomas Schuster \orcidlink{0000-0002-9539-1627} \\
    Hochschule Pforzheim \\
    \texttt{thomas.schuster@hs-pforzheim.de}
    \And
    Nico Döring \orcidlink{0009-0004-2018-5417} \\
    XPACE GmbH \\
    \texttt{nico.doering@xpace.de}
    \And 
    Robin Krüger \orcidlink{0009-0004-9937-7437} \\
    XPACE GmbH \\
    \texttt{robin.krueger@xpace.de}}

\date{Februar 2025}
\begin{document}

\maketitle

\begin{abstract}
Die Entwicklung von Large Language Models (LLMs) hat bedeutende Fortschritte in der Verarbeitung natürlicher Sprache ermöglicht und zahlreiche Anwendungen in verschiedenen Branchen hervorgebracht. Viele LLM-basierte Lösungen werden jedoch als offene Systeme betrieben, die auf Cloud-Dienste angewiesen sind und Risiken für die Vertraulichkeit und Sicherheit von Daten mit sich bringen. Um diesen Herausforderungen zu begegnen, benötigen Organisationen geschlossene LLM-Systeme, die Datenschutzvorgaben einhalten und gleichzeitig eine hohe Leistungsfähigkeit gewährleisten. In diesem Artikel stellen wir eine Referenzarchitektur für die Entwicklung geschlossener, LLM-basierter Systeme unter Verwendung von Open-Source-Technologien vor. Die Architektur bietet eine flexible und transparente Lösung, die strenge Anforderungen an Datenschutz und Sicherheit erfüllt. Wir analysieren die zentralen Herausforderungen bei der Implementierung solcher Systeme, darunter Rechenkapazitäten, Datenverwaltung, Skalierbarkeit und Sicherheitsrisiken. Zudem präsentieren wir eine Evaluationspipeline, die eine systematische Bewertung der Systemleistung und Compliance ermöglicht.
\end{abstract}


\section{Einleitung}
Die rasante Entwicklung von Large Language Models (LLMs) hat in den letzten Jahren erhebliche Fortschritte in der natürlichen Sprachverarbeitung ermöglicht und zahlreiche innovative Anwendungen in verschiedenen Branchen hervorgebracht. Diese Modelle bieten leistungsfähige KI-gestützte Lösungen, die von automatisierter Textgenerierung bis hin zur Entscheidungsunterstützung reichen. Dennoch wirft ihr Einsatz bedeutende Herausforderungen im Hinblick auf Datenschutz und Sicherheit auf.

Derzeit werden LLM-basierte Softwarelösungen überwiegend als \emph{offene Systeme} betrieben. Ein \textbf{offenes System} ist über das Internet zugänglich und nutzt externe Ressourcen oder Cloud-Dienste für die Datenverarbeitung. Dies bedeutet jedoch, dass sensible Informationen an Dritte übermittelt werden, wodurch Risiken hinsichtlich Vertraulichkeit, Integrität und Datenschutz entstehen. Unternehmen und Organisationen, die mit schutzbedürftigen Daten arbeiten, stehen daher vor der Herausforderung, alternative Betriebsmodelle zu etablieren, die eine sichere und datenschutzkonforme Nutzung von LLMs ermöglichen.

Ein \textbf{geschlossenes System} stellt eine vielversprechende Alternative dar. Hierbei werden alle Datenverarbeitungsprozesse innerhalb der eigenen Infrastruktur durchgeführt, ohne dass Informationen an externe Server übermittelt werden. Dies ermöglicht eine vollständige Kontrolle über den Datenfluss und verbessert die Einhaltung regulatorischer Vorgaben, insbesondere in sensiblen Bereichen wie dem Gesundheitswesen, der Finanzbranche oder der öffentlichen Verwaltung.

Obwohl geschlossene LLM-Systeme erhebliche Vorteile in Bezug auf Datenschutz und Sicherheit bieten, ist ihre Entwicklung mit erheblichen technischen und organisatorischen Herausforderungen verbunden. Dazu zählen hohe Anforderungen an Rechenressourcen, effiziente Datenverarbeitung, die Integration von Open-Source-Technologien sowie die Sicherstellung von Transparenz und Flexibilität.

Trotz der steigenden Nachfrage fehlt es bislang an etablierten Referenzmodellen und praxisnahen Leitlinien für die Entwicklung solcher Systeme. Entwickler und Entscheidungsträger stehen daher vor der Aufgabe, maßgeschneiderte Architekturen zu konzipieren, die sowohl leistungsfähig als auch datenschutzkonform sind.

Diese Publikation stellt eine Referenzarchitektur für die Implementierung eines geschlossenen, LLM-basierten Systems vor, das auf Open-Source-Technologien basiert. Ziel ist es, eine flexible und skalierbare Lösung bereitzustellen, die den hohen Anforderungen an Datenschutz und Datensicherheit gerecht wird. Neben technischen Aspekten werden auch organisatorische Rahmenbedingungen betrachtet, um ein umfassendes Verständnis der Herausforderungen und Lösungsansätze zu vermitteln.

Der Aufbau dieser Publikation ist wie folgt strukturiert: Kapitel~\ref{sec:motivation} erläutert die zentralen Motivationen für die Entwicklung geschlossener LLM-Systeme und geht dabei auf Datenschutz, Datensicherheit, Transparenz, Flexibilität und die Vermeidung von Vendor Lock-in ein. Kapitel~\ref{sec:challenges} analysiert die wesentlichen Herausforderungen bei der Umsetzung solcher Systeme. In Kapitel~\ref{sec:architecture} wird die vorgeschlagene Referenzarchitektur detailliert beschrieben, gefolgt von einer Diskussion in Kapitel~\ref{sec:discussion}. Abschließend fasst Kapitel~\ref{sec:conclusion} die Ergebnisse zusammen und gibt einen Ausblick auf zukünftige Entwicklungen.

Durch die Bereitstellung einer umfassenden Referenzarchitektur leistet diese Arbeit einen Beitrag zur Förderung geschlossener LLM-Systeme und zur Bewältigung der aktuellen Herausforderungen im Bereich Datenschutz und Datensicherheit.

\section{Motivation}
\label{sec:motivation}

Die Implementierung von LLMs in geschlossenen Systemen gewinnt zunehmend an Bedeutung, insbesondere in Anwendungsbereichen, in denen Datenschutz und Datensicherheit höchste Priorität haben. Während offene LLM-Plattformen eine schnelle und einfache Integration ermöglichen, stehen sie oft im Widerspruch zu den strengen datenschutzrechtlichen Anforderungen, die in vielen Branchen gelten. Dies betrifft insbesondere Bereiche wie das Gesundheitswesen, den Finanzsektor oder die öffentliche Verwaltung, in denen personenbezogene oder kritische Daten verarbeitet werden.

Ein prägnantes Beispiel hierfür ist der Einsatz von LLMs in der medizinischen Diagnostik. Während KI-gestützte Sprachmodelle in der Lage sind, umfangreiche medizinische Literatur auszuwerten und Diagnosevorschläge zu generieren, stellt die Verarbeitung sensibler Patientendaten in offenen Systemen ein erhebliches Risiko dar. Die Nutzung eines geschlossenen Systems ermöglicht es, sämtliche Verarbeitungsschritte innerhalb der kontrollierten Infrastruktur eines Krankenhauses oder einer medizinischen Einrichtung durchzuführen, wodurch die Einhaltung regulatorischer Vorgaben wie der Datenschutz-Grundverordnung (DSGVO) sichergestellt werden kann.

In diesem Kapitel werden die zentralen Motivationsfaktoren für die Entwicklung eines LLM-basierten geschlossenen Systems strukturiert dargestellt. Zunächst erörtern wir die die Bedeutung des Datenschutzes und die Notwendigkeit, personenbezogene Daten vor unbefugtem Zugriff zu schützen (siehe~\ref{sec:datenschutz}). In Abschnitt~\ref{sec:datensicherheit} betrachten wir Sicherheitsaspekte, insbesondere hinsichtlich potenzieller Angriffe und Schutzmechanismen und gehen anschließend auf die Notwendigkeit einer nachvollziehbaren und erklärbaren Modellarchitektur ein (siehe~\ref{sec:transparenz}), um regulatorischen und ethischen Anforderungen gerecht zu werden. In Abschnitt~\ref{sec:flexibilitaet_vendorlockin} diskutieren wir die Anpassungsfähigkeit geschlossener Systeme an unternehmensspezifische Anforderungen sowie die Vermeidung von Vendor Lock-in durch den Einsatz offener und erweiterbarer Technologien. Schließlich wird in Abschnitt~\ref{sec:wettbewerb} erörtert, wie Unternehmen durch die Individualisierung geschlossener LLM-Systeme Wettbewerbsvorteile erzielen können.

\subsection{Datenschutz}
\label{sec:datenschutz}

Der Schutz personenbezogener Daten ist in der heutigen digitalen Welt von zentraler Bedeutung. Mit der Einführung der Datenschutz-Grundverordnung (DSGVO) in der Europäischen Union wurden die Anforderungen an den Umgang mit sensiblen Daten erheblich verschärft. Offene LLM-Systeme, die Daten an externe Server senden oder in der Cloud verarbeiten, bergen erhebliche Risiken in Bezug auf Datenschutzverletzungen, unautorisierte Zugriffe und potenzielle Datenweitergaben an Dritte. Unternehmen, die mit vertraulichen Informationen arbeiten – etwa im Gesundheitswesen, im Finanzsektor oder in der Rechtsberatung – müssen sicherstellen, dass ihre Datenverarbeitung höchsten Sicherheitsstandards entspricht.

Ein geschlossenes LLM-System ermöglicht es, sämtliche Datenverarbeitungsprozesse vollständig innerhalb der eigenen Infrastruktur durchzuführen. Dadurch wird das Risiko von Datenlecks erheblich reduziert und die Einhaltung gesetzlicher Datenschutzvorgaben sichergestellt. Unternehmen behalten so die vollständige Kontrolle darüber, welche Daten für das Training und die Inferenz genutzt werden. Darüber hinaus kann das \enquote{Recht auf Vergessenwerden} gemäß DSGVO effizienter umgesetzt werden, indem Daten vor dem Training anonymisiert oder durch gezielte Löschmechanismen spezifische Informationen aus Modellen entfernt werden. Dies stärkt nicht nur den Datenschutz, sondern erhöht auch die Anpassungsfähigkeit von Unternehmen und Organisationen an künftige regulatorische Vorgaben.

\subsection{Datensicherheit}
\label{sec:datensicherheit}

Neben dem Datenschutz ist auch die Datensicherheit von entscheidender Bedeutung. Cyberangriffe, Datenpannen und unautorisierte Zugriffe können nicht nur erhebliche finanzielle Schäden verursachen, sondern auch das Vertrauen von Kunden, Partnern und regulatorischen Institutionen nachhaltig beeinträchtigen. Offene Systeme sind besonders häufig Ziele von Angriffen, da sie über öffentlich zugängliche Schnittstellen verfügen und potenzielle Schwachstellen enthalten, die von Angreifern gezielt ausgenutzt werden können. Dies umfasst bei LLM, neben klassischen Angriffen, insbesondere Datenexfiltration oder Manipulation von Modellausgaben durch adversarielle Angriffe.

Die Implementierung eines geschlossenen Systems ermöglicht es, Sicherheitsmaßnahmen gezielt zu steuern und an spezifische Bedrohungsszenarien anzupassen. Unternehmen können individuelle Sicherheitsmaßnahmen definieren, umfassende Verschlüsselungstechniken einsetzen und regelmäßige Sicherheitsaudits durchführen. Darüber hinaus erlaubt die Kontrolle über die Systeminfrastruktur eine individuelle Reaktion auf potenzielle Bedrohungen sowie eine verbesserte Angriffserkennung, wenn kontinuierliches Monitoring und Anomalie-Detektion eingesetzt werden. Eine Reduzierung externer Schnittstellen minimiert zudem die potenzielle Angriffsfläche und trägt wesentlich zur Erhöhung der Systemsicherheit bei.

\subsection{Regulatorische Anforderungen und Transparenz}
\label{sec:transparenz}

Mit der zunehmenden Regulierung von KI-Systemen, insbesondere durch den EU AI Act und ähnlichen gesetzlichen Rahmenwerken, stehen Unternehmen vor der Herausforderung, den Einsatz von LLMs an strenge rechtliche Vorgaben anzupassen. In vielen Branchen, darunter das Gesundheitswesen, der Finanzsektor und die öffentliche Verwaltung, sind Unternehmen verpflichtet, den Einsatz von KI-Systemen nachvollziehbar und regelkonform zu gestalten. Geschlossene Systeme bieten hier entscheidende Vorteile, da sie eine bessere Kontrolle über Datenflüsse und das Modellverhalten ermöglichen.

Transparenz in der Funktionsweise LLM-basierter Systeme ist somit in vielen Bereichen von zentraler Bedeutung. Besonders in regulierten Branchen müssen die Entscheidungsprozesse von KI-Systemen nachvollziehbar und erklärbar sein, um regulatorische Anforderungen zu erfüllen und auch um das Vertrauen der Nutzer zu stärken. Offene Systeme bieten jedoch oft nur begrenzte Einblicke in die internen Mechanismen der Modelle, was zu Compliance-Problemen und Unsicherheiten hinsichtlich der Modellentscheidungen führen kann.

Ein geschlossenes System ermöglicht hingegen den vollständigen Zugriff auf Quellcode, Modellparameter und Trainingsdaten. Dies erleichtert nicht nur die Überprüfung und Validierung der Modelle durch interne und externe Prüfinstanzen, sondern erlaubt auch eine gezielte Anpassung an spezifische Anforderungen und ethische Leitlinien. Durch den Einsatz von Open-Source-Tools kann zudem sichergestellt werden, dass keine versteckten Funktionen, ungewollten Datenweitergaben oder Sicherheitslücken im System vorhanden sind. Ergänzend können Auditing-Mechanismen und Protokollierungstechniken implementiert werden, um eine lückenlose Nachvollziehbarkeit aller Modellentscheidungen zu gewährleisten. Damit wird nicht nur die Transparenz erhöht, sondern auch das Risiko von unbeabsichtigten Verzerrungen oder unerklärbaren Entscheidungen reduziert.

Durch die verbesserte Transparenz können Unternehmen auch die Compliance sichertellen. Die Einhaltung von Vorgaben ist zudem nicht nur aus rechtlicher Sicht von Bedeutung, sondern trägt auch maßgeblich zur Stärkung des Vertrauens von Kunden und Partnern bei. Geschlossene Systeme ermöglichen es Unternehmen, nachzuweisen, dass sie verantwortungsvoll mit sensiblen Daten umgehen und sich an geltende Vorschriften sowie selbst definierte, strengere Vorgaben halten. Dies kann auch in der Zusammenarbeit mit Behörden und regulatorischen Institutionen von Vorteil sein, da Unternehmen ihre KI-gestützten Entscheidungsprozesse detailliert dokumentieren und auditierbar machen können. Durch die Kombination von Transparenz und strikter Compliance können geschlossene LLM-Systeme so maßgeblich zur sicheren und regelkonformen Nutzung von KI-Technologien beitragen.

\subsection{Flexibilität und Vermeidung von Vendor Lock-in}
\label{sec:flexibilitaet_vendorlockin}

Ein weiterer entscheidender Faktor bei der Wahl eines LLM-Systems ist die Flexibilität, das System an spezifische Anforderungen anzupassen, sowie die Vermeidung von Vendor Lock-in. Offene LLM-Plattformen bieten oftmals Lösungen, die nicht immer den individuellen Bedürfnissen eines Unternehmens entsprechen. Ein geschlossenes System ermöglicht es hingegen, Modelle und Prozesse nach Bedarf zu modifizieren und somit besser auf unternehmensspezifische Anforderungen einzugehen.

Unternehmen können beispielsweise individuelle Daten für das Training verwenden, domänenspezifische Sprachmodelle implementieren oder zusätzliche Funktionen integrieren. Diese Anpassungsfähigkeit schafft Wettbewerbsvorteile, da die entwickelten KI-Lösungen optimal auf die jeweiligen Geschäftsprozesse zugeschnitten sind. Wie bereits erwähnt, wird dadurch zudem eine präzisere Steuerung von Datenschutz- und Sicherheitsmaßnahmen ermöglicht, indem sensible Daten ausschließlich innerhalb der eigenen Infrastruktur verarbeitet werden.

Gleichzeitig birgt der Einsatz offener Systeme auf Basis von proprietären Diensten das Risiko eines Vendor Lock-ins, sodass Unternehmen von einzelnen Anbietern abhängig werden. Dies kann zu erhöhten Kosten, eingeschränkter Flexibilität und Schwierigkeiten beim Wechsel zu alternativen Lösungen führen. Durch die Nutzung von Open-Source-Tools und die Nutzung oder Entwicklung eines geschlossenen Systems können Unternehmen dieses Risiko minimieren. Sie behalten die Kontrolle über die Technologie und können bei Bedarf Anpassungen vornehmen oder auf alternative Lösungen umsteigen, ohne an einen bestimmten Anbieter gebunden zu sein. Dies fördert nicht nur die technologische Unabhängigkeit, sondern kann auch langfristig Kosten einsparen und die Wettbewerbsfähigkeit steigern.

\subsection{Wettbewerbsvorteile durch Individualisierung}
\label{sec:wettbewerb}

Die Möglichkeit, LLM-Systeme individuell anzupassen, stellt für Unternehmen zudem einen entscheidenden Wettbewerbsvorteil dar. Ein geschlossenes System erlaubt es, maßgeschneiderte Funktionen und Services zu entwickeln, die spezifisch auf die Anforderungen der jeweiligen Branche oder Kunden zugeschnitten sind. Dies kann zu einer stärkeren Kundenbindung und einer verbesserten Marktposition führen, da Unternehmen in der Lage sind, differenzierte und optimierte Lösungen anzubieten, die sich von Standardlösungen abheben.

Darüber hinaus trägt die individuelle Anpassung dazu bei, interne Prozesse effizienter zu gestalten und Arbeitsabläufe zu optimieren. Beispielsweise können branchenspezifische Fachbegriffe und unternehmensspezifische Formate direkt in die Modelle integriert werden, wodurch die Genauigkeit und Relevanz der Ergebnisse gesteigert wird. Dies führt nicht nur zu Kosteneinsparungen, sondern auch zu einer höheren Produktivität, da sich manuelle Nachbearbeitungen reduzieren lassen. 

Zudem ermöglicht eine anpassbare Architektur die nahtlose Integration in bestehende IT-Systeme, wodurch Unternehmen ihre digitalen Prozesse weiter automatisieren und optimieren können. Die Möglichkeit, eigene Datenquellen und spezialisierte Algorithmen in das System einzubinden, stellt sicher, dass das LLM den spezifischen Anforderungen gerecht wird und langfristig zur Effizienzsteigerung beiträgt. Insgesamt erlaubt die Individualisierung eines geschlossenen LLM-Systems eine nachhaltige Differenzierung gegenüber der Konkurrenz und steigert die Wettbewerbsfähigkeit in datenintensiven Märkten.

\section{Herausforderungen}
\label{sec:challenges}

Die Entwicklung eines geschlossenen LLM-Systems stellt eine vielschichtige Aufgabe dar, die eine Vielzahl von Herausforderungen mit sich bringt. Sowohl technische als auch organisatorische Herausforderungen müssen beachtet werden, um ein System zu realisieren, das den hohen Anforderungen an Datenschutz, Datensicherheit, Effizienz und Funktionalität gerecht wird. 

In diesem Kapitel werden die wesentlichen Herausforderungen in drei Hauptkategorien unterteilt: technische Herausforderungen (siehe~\ref{sec:technische_herausforderungen}), regulatorische und ethische (siehe~\ref{sec:regulatorische_herausforderungen}) sowie organisatorische und betriebswirtschaftliche Herausforderungen (siehe~\ref{sec:organisatorische_herausforderungen}).

\subsection{Technische Herausforderungen}
\label{sec:technische_herausforderungen}

Die technische Umsetzung eines geschlossenen LLM-Systems erfordert eine präzise Planung zur effizienten Nutzung der verfügbaren Ressourcen. Dabei lassen sich folgende zentrale technische Herausforderungen identifizieren.

\paragraph{Rechenleistung und Modellkomplexität}

LLMs sind hochkomplexe Modelle, die beträchtliche Rechenressourcen benötigen - auch wenn der Ressourcenverbrauch zunehmend Beachtung findet und Modelle inzwischen effizienter betrieben werden können. Sowohl das Training und die Inferenz erfordern spezialisierte Hardware wie Grafikprozessoren (Graphics Processing Units, kurz \emph{GPU}) oder Tensor-Prozessoren (Tensor Processing Units, kurz \emph{TPU}). In geschlossenen Systemen stehen jedoch oft nicht die gleichen skalierbaren Infrastrukturen zur Verfügung wie in offenen Systemen und öffentlichen Cloud-Umgebungen. Dies macht es notwendig, Modelle für lokale Umgebungen zu optimieren – etwa durch Quantisierung, Modellkomprimierung oder den Einsatz spezialisierter Small Language Models (\emph{SLMs}), die mittels effizienten Methoden (bspw. Parameter-efficient Finetuning, kurz \emph{PEFT}) angepasst werden können.

\paragraph{Skalierbarkeit und Performance}

Die Fähigkeit eines Systems, Lastspitzen zu bewältigen und flexibel zu skalieren, ist entscheidend für den praktischen Einsatz. In geschlossenen Umgebungen sind die Möglichkeiten zur Skalierung häufig begrenzt. Die Implementierung effizienter Algorithmen, paralleler Verarbeitungsstrategien und Caching-Techniken ist daher essenziell, um kurze Antwortzeiten und stabile Performance zu gewährleisten.

\paragraph{Integration und Kompatibilität}

Die Einbindung eines geschlossenen LLM-Systems in bestehende IT-Infrastrukturen stellt eine weitere Herausforderung dar. Eine nahtlose Integration setzt standardisierte Schnittstellen (Application Programming Interface, kurz \emph{API}), kompatible Datenformate und eine sorgfältige Abstimmung zwischen verschiedenen Systemkomponenten voraus. Middleware-Lösungen können helfen, Interoperabilitätsprobleme zu minimieren und eine reibungslose Einbettung in bestehende Prozesse (Workflows) zu ermöglichen.

\paragraph{Datenmanagement}

Die Verarbeitung großer Datenmengen erfordert effiziente Speicher- und Abrufmechanismen. Der Einsatz von Vektordatenbanken für semantische Suchanfragen und NoSQL-Datenbanken für unstrukturierte Daten kann zur Leistungssteigerung beitragen. Gleichzeitig ist eine durchgehende Sicherstellung der Datenqualität entscheidend, um Verzerrungen und Modellfehler zu vermeiden.

\paragraph{Zukunftssicherheit und Weiterentwicklung}

Da sich die KI-Technologie rasant weiterentwickelt, muss ein geschlossenes LLM-System langfristig anpassbar bleiben. Die Architektur des Systems sollte so gestaltet sein, dass neue Modelle und Technologien unkompliziert integriert werden können. Auch regulatorische Vorgaben können sich ändern, sodass eine flexible Systemarchitektur essenziell ist, um langfristige Compliance sicherzustellen.

\subsection{Regulatorische und ethische Herausforderungen}
\label{sec:regulatorische_herausforderungen}

Der Schutz sensibler Daten ist allgemein eine der größten Herausforderungen und auch bei geschlossenenen LLM-Systemen besonders zu beachten. Hierbei sind nicht nur klassische Sicherheitsmaßnahmen erforderlich, sondern auch gezielte Schutzmechanismen gegen LLM-spezifische Bedrohungen wie Prompt Injection, Data Poisoning, Model Inversion Attacks oder adversarielle Angriffe.

\paragraph{Datenschutz und Einhaltung gesetzlicher Vorschriften}

Die Datenschutz-Grundverordnung (DSGVO) und weitere regulatorische Vorgaben erfordern eine konsequente Umsetzung von Datenschutzmaßnahmen. Dazu zählen die Implementierung von Anonymisierungsverfahren, ein effektives Einwilligungsmanagement sowie Mechanismen zur Löschung personenbezogener Daten nach geltenden Vorschriften.

\paragraph{Sicherheit und Schutz vor Angriffen}

Ein sicheres System muss Verschlüsselungstechniken für die Datenübertragung und -speicherung implementieren sowie strikte Zugriffskontrollen durchsetzen. Ergänzend sind regelmäßige Sicherheitsüberprüfungen und eine kontinuierliche Überwachung durch Monitoring- und Logging-Systeme notwendig, um Schwachstellen frühzeitig zu identifizieren. Zudem sind LLMs anfällig für adversarielle Angriffe wie Prompt Injection oder Data Poisoning, weshalb Sicherheitsmaßnahmen auf Modell- und Infrastruktur-Ebene erforderlich sind.

\paragraph{Ethik, Bias und Fairness}

KI-Modelle neigen dazu, bestehende Verzerrungen in Trainingsdaten zu übernehmen. Die Implementierung von Fairness-Metriken und kontinuierliche Tests auf Verzerrungen (Bias) sind notwendig, um diskriminierende Tendenzen in den Ergebnissen zu minimieren. Zusätzlich muss sichergestellt werden, dass Entscheidungsprozesse erklärbar und nachvollziehbar sind, insbesondere in kritischen Anwendungsbereichen.

\paragraph{Verantwortung und Haftung}

Die Frage nach der Verantwortlichkeit bei Fehlentscheidungen oder schädlichen Auswirkungen von KI-Systemen ist komplex. Unternehmen müssen klare Richtlinien im Umgang mit KI-gestützten Entscheidungen entwickeln, um Verantwortung zu übernehmen und Risiken zu managen. Dazu gehört die Definition von Kontrollmechanismen für den Einsatz von LLMs in sensiblen Bereichen sowie die Möglichkeit, Fehler zu korrigieren und nachvollziehbare Entscheidungsprozesse sicherzustellen.

\subsection{Organisatorische und betriebswirtschaftliche Herausforderungen}
\label{sec:organisatorische_herausforderungen}
Neben den technischen und regulatorischen Herausforderungen spielen auch organisatorische und wirtschaftliche Faktoren eine zentrale Rolle bei der erfolgreichen Umsetzung eines geschlossenen LLM-Systems.

\paragraph{Interdisziplinäre Zusammenarbeit}
Die erfolgreiche Umsetzung erfordert die Zusammenarbeit von Expert:innen, besonders aus den Bereichen Informatik, Recht und Ethik. Die Koordination solcher Teams stellt eine Herausforderung dar und erfordert effektive Management- und Kommunikationsstrategien.

\paragraph{Kompetenzaufbau und Schulung}
Der Betrieb eines geschlossenen LLM-Systems erfordert spezialisiertes Wissen. Unternehmen müssen in Schulungsmaßnahmen investieren, um interne Kompetenzen für die Wartung, Optimierung und Absicherung der Systeme aufzubauen.

\paragraph{Akzeptanz und Veränderungsmanagement}
Neue Technologien stoßen oft auf Vorbehalte. Um die Akzeptanz innerhalb des Unternehmens zu fördern, sind transparente Kommunikationsstrategien und ein strukturiertes Veränderungsmanagement essenziell.

\paragraph{Kosten- und Ressourcenmanagement}

Die finanziellen und operativen Ressourcen müssen effizient geplant werden, um die Umsetzung eines geschlossenen LLM-Systems wirtschaftlich tragfähig zu gestalten. Die Anschaffung und der Betrieb der benötigten Hardware sind mit erheblichen Kosten verbunden. Unternehmen sollten daher eine fundierte Kosten-Nutzen-Analyse durchführen und gegebenenfalls alternative Finanzierungsmodelle prüfen. Zudem ist eine durchdachte Planung der Ressourcennutzung – sowohl personell als auch materiell – essenziell, um den Betrieb nachhaltig zu gewährleisten und Engpässe zu vermeiden.

\section{Referenzarchitektur}
\label{sec:architecture}

Die Entwicklung eines datenschutzkonformen und modularen LLM-Systems erfordert eine durchdachte Architektur, die sowohl technische Anforderungen als auch Sicherheitsrichtlinien berücksichtigt. In diesem Kapitel wird eine detaillierte Referenzarchitektur vorgestellt, die auf Open-Source-Technologien basiert und eine skalierbare, leistungsstarke und sichere Lösung für verschiedene Anwendungsfälle bietet. Die Architektur gliedert sich in mehrere Schichten und Module, die nahtlos zusammenarbeiten, um ein flexibles Gesamtsystem zu bilden.

\subsection{Überblick}

Die vorgeschlagene Architektur besteht aus zwei übergeordneten Kategorien: \textbf{technische Kernkomponenten} und \textbf{architekturübergreifende Querschnittsfunktionen}.

\textbf{Technische Kernkomponenten:}
\begin{itemize}
    \item \textbf{Frontend}: Benutzeroberfläche für die Interaktion mit dem System.
    \item \textbf{Backend}: Geschäftslogik und Kommunikationszentrale zwischen den Komponenten.
    \item \textbf{LLM-Inferenz-Engine}: Ausführung der Sprachmodelle und Embedding-Berechnungen.
    \item \textbf{Caching}: Performance-Optimierung durch Zwischenspeicherung.
    \item \textbf{Datenbanken}: Speicherung und Verwaltung strukturierter und unstrukturierter Daten.
\end{itemize}

\textbf{Architekturübergreifende Querschnittsfunktionen:}
\begin{itemize}
    \item \textbf{Testing und Evaluation}: Qualitätssicherung und Leistungsbewertung.
    \item \textbf{Sicherheit}: Maßnahmen zum Schutz von Daten und Systemintegrität.
    \item \textbf{Monitoring und Logging}: Überwachung des Systems und Protokollierung.
    \item \textbf{Deployment und Orchestrierung}: Bereitstellung und Skalierung des Systems.
\end{itemize}

\begin{figure}
    \centering
    \includegraphics[width=0.8\linewidth]{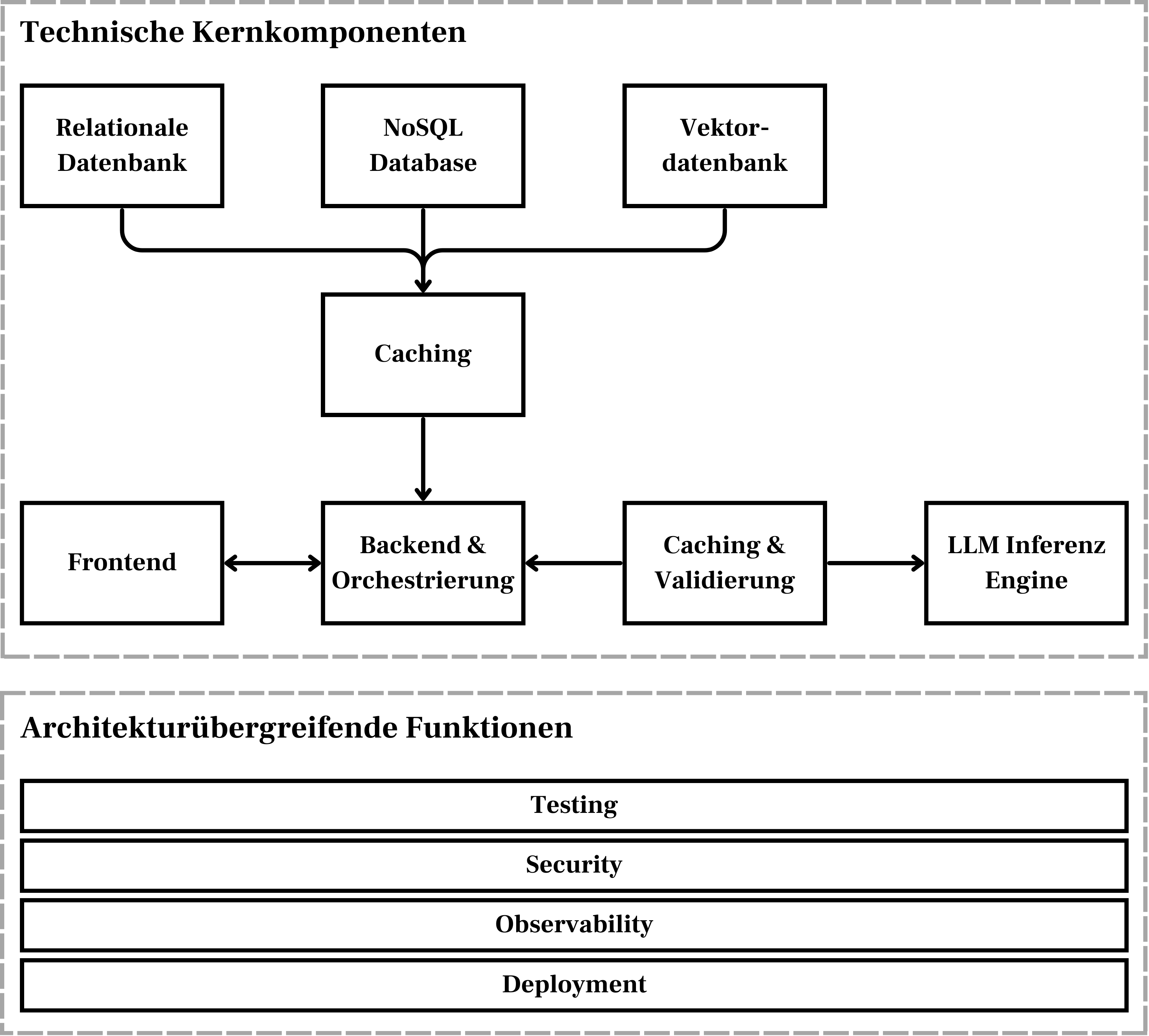}
    \caption{Referenzarchitektur eines geschlossenen LLM-Systems}
    \label{fig:architecture}
\end{figure}

Abbildung~\ref{fig:architecture} zeigt eine schematische Darstellung der Architektur und die Interaktion zwischen den einzelnen Komponenten.

\subsection{Komponenten}
Im Folgenden werden die einzelnen Komponenten im Detail erläutert und es wird aufgezeigt, wie sie zusammenwirken, um ein effizientes und sicheres LLM-System zu realisieren.

\subsubsection{Frontend}

Das Frontend bildet die zentrale Schnittstelle zwischen dem Benutzer und dem System und ermöglicht eine intuitive, effiziente und barrierefreie Interaktion. Ein gut gestaltetes Frontend sollte sich responsiv verhalten, sodass es sich automatisch an verschiedene Geräte und Bildschirmgrößen anpasst, und gleichzeitig benutzerfreundlich sein – dies umfasst eine intuitive Bedienung, ein ansprechendes Design sowie eine konsistente Nutzerführung.

Ebenso entscheidend sind leistungsoptimierte Ladezeiten und eine flüssige Interaktion, um die Nutzererfahrung nicht durch Verzögerungen zu beeinträchtigen. Dazu gehört auch ein effektives Fehlermanagement, das dem Nutzer verständliche Rückmeldungen bei Eingabefehlern oder Systemproblemen liefert.

Aus Sicherheitsgründen muss das Frontend verlässliche Authentifizierungs- und Autorisierungsmechanismen implementieren, um sicherzustellen, dass Nutzer ausschließlich auf die für sie bestimmten Inhalte und Funktionen zugreifen können. Datenschutz und sichere Datenübertragung – etwa durch die Nutzung von HTTPS und der Vermeidung von Sicherheitslücken wie Cross-Site Scripting – sind dabei unerlässlich.

Ein weiterer wichtiger Aspekt ist die Barrierefreiheit: Das Frontend sollte so gestaltet sein, dass auch Menschen mit Einschränkungen es problemlos nutzen können. Internationalisierung und Lokalisierung spielen ebenfalls eine zentrale Rolle, um die Anwendung für ein globales Publikum verständlich und kulturell angemessen bereitzustellen.

Nicht zuletzt sollte das Frontend skalierbar und wartungsfreundlich aufgebaut sein. Dies wird durch modulare Komponenten, gut dokumentierten Code und die Nutzung zeitgemäßer Frameworks und Tools erreicht, um auch künftige Erweiterungen problemlos integrieren zu können.

Im Open-Source-Ökosystem gibt es eine Vielzahl von Frameworks und Bibliotheken, die zur Frontend-Entwicklung eingesetzt werden können. Besonders beliebt sind dabei die folgenden:

\begin{itemize}
\item \textbf{React} mit \textbf{Next.js} ~\cite{react, nextjs}: Ein weit verbreitetes JavaScript-Framework für den Aufbau dynamischer Benutzeroberflächen, das mit Next.js serverseitiges Rendering, statische Seitenerzeugung und API-Routen unterstützt, wodurch komplexe, skalierbare und suchmaschinenfreundliche Webanwendungen entstehen.
\item \textbf{Vue.js} mit \textbf{Nuxt.js}~\cite{nuxtjs}: Ein progressives JavaScript-Framework mit einem fokus auf einfache Integration und Komponentenstruktur; in Kombination mit Nuxt.js bietet es serverseitiges Rendering, Modulunterstützung und eine starke Entwicklererfahrung für schnelle Entwicklung robuster Webanwendungen.
\item \textbf{Svelte}~\cite{svelte}: Ein kompiliertes Frontend-Framework, das bereits zur Build-Zeit optimierten JavaScript-Code erzeugt und somit besonders performante Webanwendungen ermöglicht, ohne dass ein virtuelles DOM erforderlich ist.
\item \textbf{Streamlit}~\cite{streamlit}: Ein auf Python basiertes Framework, das die schnelle Entwicklung interaktiver Webanwendungen für Datenvisualisierung, maschinelles Lernen und Prototyping ohne tiefgreifende Frontend-Kenntnisse ermöglicht.
\item \textbf{Flutter}~\cite{flutter}: Ein von Google entwickeltes UI-Toolkit, mit dem plattformübergreifende Anwendungen aus einer einzigen Codebasis erstellt werden können, die sowohl auf mobilen Geräten als auch im Web eine konsistente Benutzererfahrung bieten.
\end{itemize}

\subsubsection{Backend}

Das Backend bildet das Herzstück des Systems, indem es die Geschäftslogik implementiert und die Kommunikation zwischen dem Frontend, den Datenbanken und der LLM-Inferenz-Engine koordiniert. Es ist maßgeblich für die Datenverarbeitung, Sicherheit, Performance und Systemstabilität verantwortlich. Ein leistungsfähiges Backend sorgt dafür, dass Anfragen effizient bearbeitet, Daten konsistent verwaltet und externe Dienste zuverlässig eingebunden werden.

Ein wesentliches Merkmal eines guten Backends ist die Bereitstellung gut strukturierter API-Endpunkte, die über Standards wie RESTful, GraphQL oder gRPC eine reibungslose Kommunikation zwischen den Systemkomponenten ermöglichen. Hierbei ist eine klare Schnittstellendefinition ebenso wichtig wie die Implementierung einer API-Versionierung, um Kompatibilität bei Weiterentwicklungen sicherzustellen.

Die Sicherheit spielt eine zentrale Rolle. Neben der Implementierung robuster Authentifizierungs- und Autorisierungsmechanismen (im Zusammenspiel mit dem Frontend) müssen Schutzmaßnahmen gegen gängige Sicherheitslücken wie SQL-Injections, Cross-Site Scripting (XSS) und Cross-Site Request Forgery (CSRF) ergriffen werden. Ebenso sollten Daten sowohl im Transit als auch im Ruhezustand verschlüsselt werden, um Datenschutz und Integrität zu gewährleisten.

Ein stabiles Backend zeichnet sich durch eine effektive Fehlerbehandlung und hohe Resilienz aus. Fehler sollten nicht nur intern erfasst, sondern dem Nutzer auch durch verständliche Fehlermeldungen kommuniziert werden. Fallback-Strategien und ein umfassendes Monitoring ermöglichen es, Probleme frühzeitig zu erkennen und zu beheben, wodurch Systemausfälle minimiert werden.

Auch die Skalierbarkeit und Performance sind entscheidend. Das Backend muss in der Lage sein, unter hoher Last stabil zu bleiben. Dies wird durch Lastverteilung, horizontale Skalierung und den Einsatz von Caching-Mechanismen wie Redis erreicht.

Ein weiterer wichtiger Aspekt ist das Datenmanagement. Das Backend sollte eine effiziente Anbindung an verschiedene Datenquellen bieten, sei es an relationale Datenbanken oder NoSQL-Lösungen. Hierbei müssen Datenkonsistenz und Integrität stets gewährleistet sein.

Neben den oben genannten Fullstack-Frameworks wie Streamlit, Next.js und Vue.js bieten sich die folgenden Open-Source-Projekt für die Backend-Entwicklung an:

\begin{itemize}
\item \textbf{FastAPI}~\cite{fastapi}: Modernes, asynchrones Python-Framework für schnelle API-Entwicklung mit automatischer Dokumentation und Typunterstützung.
\item \textbf{Django} mit \textbf{Django REST Framework}~\cite{django}: Umfassendes Python-Framework für robuste Webanwendungen mit integriertem ORM und leistungsfähigem API-Support.
\item \textbf{Flask}~\cite{flask}: Flexibles, leichtgewichtiges Python-Framework für einfache Anwendungen und Microservices.
\item \textbf{Express.js}~\cite{express}: Minimalistisches Node.js-Framework für schnelle Entwicklung von Webanwendungen und APIs.
\item \textbf{ASP.NET Core}~\cite{aspnetcore}: Plattformübergreifendes C\#-Framework von Microsoft für performante und skalierbare Webanwendungen.
\item \textbf{Spring Boot}~\cite{springboot}: Java-basiertes Framework für schnelle Entwicklung von Microservices mit vorkonfigurierter Infrastruktur.
\end{itemize}

Als Teil des Backends koordiniert die \textbf{KI-Orchestrierungskomponente} die Interaktion zwischen verschiedenen KI-Komponenten, Datenquellen und externen Systemen. Ziel ist es, komplexe Verarbeitungsketten effizient zu steuern und dabei Flexibilität, Zuverlässigkeit und Wartbarkeit sicherzustellen. Wichtige Kriterien sind dabei eine modulare Architektur, die den Austausch und die Kombination verschiedener Modelle, Tools und Datenquellen ermöglicht, sowie Erweiterbarkeit, um neue Komponenten ohne großen Aufwand zu integrieren.

Da Orchestrierung oft mehrere Verarbeitungsschritte (z. B. Vorverarbeitung, Modellaufrufe, Nachbearbeitung) umfasst, sind Workflow-Management und Fehlerbehandlung entscheidend. Ein gutes Orchestrierungs-Backend stellt sicher, dass Prozesse robust ablaufen, bei Fehlern wiederaufgesetzt werden können und Fallback-Mechanismen vorhanden sind. Die Latenzoptimierung bleibt auch hier relevant, insbesondere bei Abhängigkeiten zwischen Verarbeitungsschritten.

Transparenz und Nachvollziehbarkeit sind weitere zentrale Aspekte: Logging, Tracing und Monitoring helfen dabei, Datenflüsse zu überwachen und Engpässe oder Fehler schnell zu identifizieren. Zudem sollte das Backend Datenkonsistenz wahren, vor allem wenn Informationen aus verschiedenen Quellen kombiniert werden.

Für die Orchestrierung von LLMs und Verarbeitungspipelines können folgende Tools integriert werden:

\begin{itemize}
\item \textbf{LangChain}~\cite{langchain}: Ein Framework zur Erstellung komplexer KI-gestützter Anwendungen, das die Verkettung von LLM-Aufrufen, Integration externer Datenquellen, Verwaltung von Konversationen und Workflow-Orchestrierung ermöglicht.
\item \textbf{Haystack}~\cite{haystack}: Ein leistungsfähiges Open-Source-Framework für die Entwicklung von Frage-Antwort-Systemen und Retrieval-Augmented Generation (RAG)-Anwendungen, das modulare Pipelines zur Verarbeitung unstrukturierter und strukturierter Daten unterstützt.
\item \textbf{Semantic Kernel}~\cite{semantickernel}: Ein von Microsoft entwickeltes Framework zur Integration von LLMs in Anwendungen, das Funktionen wie Planung, Gedächtnisverwaltung und die Orchestrierung von KI-Skills bietet, um komplexe Arbeitsabläufe zu automatisieren.
\item \textbf{LlamaIndex}~\cite{llamaindex}: Ein Tool zur effizienten Erstellung und Verwaltung von Indizes für Large Language Models (LLMs), das eine nahtlose Anbindung an verschiedene Datenquellen und eine optimierte Suche innerhalb großer Datenmengen ermöglicht.
\end{itemize}

\subsubsection{LLM-Inferenz-Engine}

Die LLM-Inferenz-Engine ist für die effiziente und zuverlässige Ausführung von Sprachmodellen verantwortlich und spielt eine zentrale Rolle bei der Verarbeitung von Benutzereingaben, der Generierung von Antworten und der Koordination der Modellaufrufe.

Ein wesentliches Kriterium dabei ist die Performance, da geringe Latenzzeiten und ein hoher Durchsatz entscheidend für eine gute Nutzererfahrung sind. Insbesondere bei Echtzeitanwendungen muss die Inferenz-Engine schnelle Antwortzeiten gewährleisten, ohne dabei die Qualität der generierten Texte zu beeinträchtigen. Durch Batching-Mechanismen und optimierte Hardware-Nutzung kann die Verarbeitungsgeschwindigkeit erheblich gesteigert werden.  

Die Skalierbarkeit ist ein weiterer entscheidender Faktor, um auch bei hoher Last oder komplexen Aufgaben stabil zu bleiben. Eine gute Inferenz-Engine unterstützt den Betrieb auf mehreren GPUs sowie in verteilten Umgebungen und kann je nach Bedarf automatisch Ressourcen skalieren. Dies ist besonders relevant, wenn mehrere parallele Anfragen verarbeitet oder große Modelle eingesetzt werden.  

Ebenso wichtig ist die Flexibilität. Die Inferenz-Engine sollte die einfache Integration neuer Modelle ermöglichen und sich problemlos an spezifische Anforderungen anpassen lassen. Dazu gehören die Unterstützung verschiedener LLM-Architekturen, die Konfiguration von Inferenzparametern (wie Temperatur oder Top-k) sowie die Möglichkeit, Modelle für spezielle Anwendungsfälle zu modifizieren oder zu kombinieren.  

Darüber hinaus spielen auch Sicherheit und Zuverlässigkeit eine wichtige Rolle: Die Engine muss sensible Daten während der Verarbeitung schützen und eine robuste Fehlerbehandlung sowie Fallback-Mechanismen bereitstellen. Ergänzend dazu sollten Monitoring-, Logging- und Tracing-Funktionen integriert sein, um die Modellnutzung, Latenzen und Fehlerraten kontinuierlich zu überwachen.  

Die folgenden Open-Source-Inferenz-Engines bieten sich für den Einsatz in geschlossenen Systemen an:

\begin{itemize}
\item \textbf{vLLM}~\cite{vllm}: Eine hochperformante Inferenz-Engine für Large Language Models, die durch effizientes Speicher-Management und optimiertes Batching extrem niedrige Latenzen bei hoher Anfragefrequenz ermöglicht und speziell für die Nutzung auf GPUs optimiert ist.
\item \textbf{Hugging Face Text Generation Inference}~\cite{huggingface}: Ein skalierbarer und leistungsfähiger Inferenzserver für Textgenerierungsmodelle, der Multi-GPU-Unterstützung, Modell-Parallelisierung und einfache Integration in Produktionsumgebungen bietet.
\item \textbf{ONNX Runtime}~\cite{onnx}: Eine plattformübergreifende Inferenz-Lösung für Modelle im ONNX-Format, die hardwareoptimierte Ausführung auf verschiedenen Geräten (CPU, GPU, FPGA) ermöglicht und für maximale Performance und Portabilität sorgt.
\item \textbf{OpenVINO}~\cite{openvino}: Ein von Intel entwickeltes Toolkit zur Optimierung und Beschleunigung der Inferenz auf Intel-Hardware (CPU, GPU, VPU), das speziell auf Edge- und Cloud-Szenarien ausgelegt ist und effiziente Nutzung vorhandener Ressourcen ermöglicht.
\item \textbf{ollama}~\cite{ollama}: Eine benutzerfreundliche Inferenzlösung für lokale Ausführung von LLMs, die einfache Modellverwaltung, schnelle Inferenzzeiten und Kompatibilität mit verschiedenen Modellarchitekturen bietet, ohne auf eine Cloud-Infrastruktur angewiesen zu sein.
\item \textbf{llama.cpp}~\cite{llamacpp}: Eine leichtgewichtige, C++-basierte Inferenz-Engine für LLaMA-Modelle, die für CPU-basierte Inferenz optimiert ist und den Einsatz großer Sprachmodelle auf Geräten mit begrenzten Ressourcen (wie Laptops oder mobilen Geräten) ermöglicht.
\end{itemize}

\subsubsection{Datenbanken}

Die effiziente Speicherung und Verwaltung von Daten ist von zentraler Bedeutung für die Leistungsfähigkeit, Skalierbarkeit und Zuverlässigkeit des Systems. Die Wahl der passenden Datenbank hängt dabei von verschiedenen Kriterien ab, die sich an den spezifischen Anforderungen der Anwendung orientieren. Wichtige Faktoren sind die Art und Struktur der Daten (z.\,B. strukturierte, semi-strukturierte oder unstrukturierte Daten), die benötigten Abfrage- und Verarbeitungsfunktionen sowie die Skalierbarkeit in Bezug auf Datenvolumen und Zugriffszahlen. Auch die Leistung spielt eine entscheidende Rolle: Eine niedrige Latenz, hoher Durchsatz und effiziente Indexierungsmechanismen sind entscheidend, um schnelle Lese- und Schreibzugriffe zu gewährleisten. 

Darüber hinaus sind Verfügbarkeit und Zuverlässigkeit essenziell, weshalb Datenbanken Funktionen wie Replikation, automatische Failover-Mechanismen und effektive Backup- und Recovery-Strategien unterstützen sollten. Die Konsistenzanforderungen müssen je nach Anwendungsfall berücksichtigt werden: Während ACID-Konformität bei transaktionskritischen Anwendungen unerlässlich ist, können bei verteilten Systemen mit hoher Verfügbarkeit eventuelle Konsistenzmodelle ausreichend sein. Auch die Sicherheit darf nicht vernachlässigt werden; dazu gehören Verschlüsselung von Daten (im Transit und im Ruhezustand), Zugriffskontrollmechanismen sowie die Einhaltung von Datenschutz- und Compliance-Vorgaben. 

Ein weiteres wichtiges Kriterium ist die Flexibilität und Erweiterbarkeit der Datenbank, um auf sich ändernde Anforderungen und Datenmodelle reagieren zu können. Die Integrationsfähigkeit in bestehende Systemlandschaften sowie die Unterstützung gängiger Schnittstellen und APIs erleichtern die Einbindung in bestehende Workflows.

Es gibt eine Vielzahl von Open-Source-Datenbanken, die folgenden bieten sich besonders an:

\begin{itemize}
\item \textbf{PostgreSQL}~\cite{postgresql}: Eine leistungsfähige relationale Datenbank für die Speicherung strukturierter Daten, die erweiterte Funktionen wie komplexe Abfragen, Transaktionen mit ACID-Konformität, JSON-Unterstützung und leistungsstarke Erweiterungen (z.\,B. PostGIS für Geodaten) bietet.
\item \textbf{MongoDB}~\cite{mongodb}: Eine dokumentenorientierte NoSQL-Datenbank für die flexible Speicherung unstrukturierter und semi-strukturierter Daten, die schemalose Datensätze, horizontale Skalierung und leistungsfähige Abfragefunktionen für JSON-ähnliche Dokumente ermöglicht.
\item \textbf{ChromaDB} und \textbf{FAISS}~\cite{chromadb, faiss}: Spezialisierte Vektordatenbanken zur effizienten Verwaltung und Suche von hochdimensionalen Embeddings, die vor allem für semantische Suche, Empfehlungssysteme und KI-gestützte Anwendungen genutzt werden; FAISS ist für maximale Geschwindigkeit auf großen Datensätzen optimiert, während ChromaDB zusätzliche Features zur Datenverwaltung bietet.
\item \textbf{Weaviate}~\cite{weaviate}: Eine skalierbare Vektordatenbank mit integrierter KI-Funktionalität, die semantische Suche, kontextbezogene Datenabfragen und die direkte Integration von Machine-Learning-Modellen ermöglicht.
\item \textbf{Redis}~\cite{redis}: Ein In-Memory-Datenspeicher, der sich sowohl als schneller Cache als auch als persistente NoSQL-Datenbank eignet, um Latenzzeiten zu minimieren und hohe Durchsatzraten zu unterstützen; bietet zudem Datenstrukturen wie Listen, Hashes und Streams.
\item \textbf{ClickHouse}~\cite{clickhouse}: Eine spaltenorientierte Datenbank, die speziell für analytische Abfragen auf großen Datenmengen optimiert ist und durch hohe Verarbeitungsgeschwindigkeit, Echtzeitanalysen sowie effiziente Komprimierung überzeugt.
\end{itemize}

\subsubsection{Caching}

Um die Systemperformance zu optimieren, ist das Caching ein entscheidender Mechanismus zur Zwischenspeicherung häufig benötigter Daten. Dadurch können Latenzzeiten erheblich reduziert, die Systemeffizienz gesteigert und gleichzeitig die Belastung der Datenquellen verringert werden.

Die Auswahl geeigneter Caching-Software ist entscheidend für die Optimierung der Systemleistung. Wichtige Kriterien sind Performance (niedrige Latenz, hoher Durchsatz) und Skalierbarkeit, sowohl vertikal als auch horizontal. Die Software sollte verschiedene Caching-Strategien (z.\,B. TTL, LRU, LFU) und Konsistenzmechanismen zur Synchronisation mit der Datenquelle unterstützen. Eine einfache Integration in bestehende Systeme, Unterstützung gängiger Schnittstellen sowie Sicherheitsfunktionen wie Zugriffskontrolle und Verschlüsselung sind essenziell. Zudem sind Monitoring-Funktionen zur Überwachung von Cache-Hit-Raten und Speicherverbrauch sowie Kostenaspekte bei der Auswahl zu berücksichtigen.

Für das Caching können die folgenden Softwarelösungen eingesetzt werden:
\begin{itemize}
\item \textbf{Redis}~\cite{redis}: Ein vielseitiger In-Memory-Datenspeicher, der als Cache, Message Broker und persistente NoSQL-Datenbank genutzt werden kann; bietet Unterstützung für verschiedene Datenstrukturen, hohe Verfügbarkeit und Replikation.
\item \textbf{Memcached}~\cite{memcached}: Ein schlanker, leistungsstarker In-Memory-Cache, der sich besonders für einfache Key-Value-Speicherung und schnelle Datenabfragen eignet; ideal zur Reduzierung von Datenbanklasten.
\item \textbf{Apache Ignite}~\cite{ignite}: Ein verteiltes In-Memory-Computing-Framework mit integriertem Caching, das hohe Performance, Datenpersistenz und Unterstützung für SQL-Abfragen sowie Machine-Learning-Workloads bietet.
\item \textbf{Hazelcast}~\cite{hazelcast}: Eine In-Memory-Datenplattform für verteilte Caching-Lösungen, die skalierbare Datenverarbeitung, Stream-Processing und einfache Integration in Cloud-Umgebungen ermöglicht.
\end{itemize}

\subsubsection{Sicherheit}

Die Sicherheit in LLM-Systemen erfordert besondere Aufmerksamkeit, da diese Systeme oft mit sensiblen Daten interagieren und spezifischen Bedrohungen ausgesetzt sind. Ein zentrales Risiko besteht in Prompt-Injection-Angriffen, bei denen Angreifer versuchen, das Modell durch manipulierte Eingaben zu unerwünschtem Verhalten zu verleiten. Um dem entgegenzuwirken, sollten Eingabevalidierungen, Filtermechanismen und Kontextbegrenzungen implementiert werden.  

Darüber hinaus ist die Vertraulichkeit der Eingabe- und Ausgabedaten entscheidend. Dies erfordert Ende-zu-Ende-Verschlüsselung, sichere Speicherung sowie Anonymisierung oder Pseudonymisierung sensibler Informationen. Um das Modell vor unbefugtem Zugriff zu schützen, sollten Zugriffskontrollen, API-Authentifizierung und Rate-Limiting eingesetzt werden.  

Ein weiteres wichtiges Thema ist die Verhinderung von Informationslecks, bei denen das Modell unbeabsichtigt vertrauliche Informationen preisgeben könnte. Hier helfen Content-Filter, Ausgabeüberwachung und das Training mit datenschutzkonformen Datensätzen.  

Um Modellmissbrauch zu verhindern, sollten Mechanismen zur Missbrauchserkennung sowie Monitoring- und Logging-Tools integriert werden. Diese sollten jedoch so konzipiert sein, dass sie keine sensiblen Daten im Klartext speichern.  

Schließlich müssen LLM-Systeme Compliance-Anforderungen erfüllen. Dies beinhaltet die Berücksichtigung von Vorschriften wie der DSGVO, regelmäßige Sicherheitsüberprüfungen und die Implementierung eines Incident-Response-Plans, um im Falle eines Vorfalls schnell reagieren zu können.

Dabei sollte insbesondere der Einsatz der folgenden Open-Source-Projekte in Betracht gezogen werden:  

\begin{itemize}
\item \textbf{Llama Guard}~\cite{llamaguard} und \textbf{Guardrails}~\cite{guardrails}: Tools zur Überwachung und Steuerung von Modellausgaben, die unerwünschte oder unsichere Inhalte erkennen und filtern, um die Sicherheit und Vertrauenswürdigkeit von LLM-Systemen zu gewährleisten.
\item \textbf{Keycloak}~\cite{keycloak}: Eine Open-Source-Lösung für Identity- und Access-Management, die Single Sign-On (SSO), rollenbasierte Zugriffskontrollen und Integrationen mit gängigen Authentifizierungsstandards wie OAuth2, OpenID Connect und SAML unterstützt.
\item \textbf{Vault} von HashiCorp~\cite{vault}: Ein Tool zur sicheren Verwaltung von Secrets, Schlüsseln und sensiblen Konfigurationsdaten, das Verschlüsselung, dynamische Geheimnisvergabe und Zugriffskontrollen bereitstellt.
\item \textbf{Open Policy Agent (OPA)}~\cite{opa}: Eine flexible Policy-Engine zur Durchsetzung von Zugriffskontrollen und Sicherheitsrichtlinien in verschiedenen Systemen und Microservices-Architekturen.
\item \textbf{Falco}~\cite{falco}: Ein Open-Source-Tool für Echtzeit-Sicherheitsüberwachung, das verdächtige Aktivitäten in Containern, Kubernetes-Umgebungen und Cloud-Infrastrukturen erkennt und Alarmierungen bereitstellt.
\end{itemize}

\subsubsection{Monitoring und Logging}

Für einen stabilen und zuverlässigen Betrieb ist eine kontinuierliche Überwachung des Systems unerlässlich. Geeignete Monitoring- und Logging-Tools ermöglichen es, den Systemzustand in Echtzeit zu überwachen, Engpässe frühzeitig zu erkennen und zeitnah auf Probleme zu reagieren. Neben der Erfassung von Systemmetriken wie CPU- und Speicherauslastung sind auch das Monitoring von Anwendungsmetriken, Logging sicherheitsrelevanter Ereignisse und die Überwachung von LLM-spezifischen Parametern (wie Inferenzzeiten und Fehlerraten) entscheidend.  

Wichtige Kriterien bei der Auswahl solcher Tools sind Skalierbarkeit, geringe Latenz, flexible Alarmierungsmechanismen und benutzerfreundliche Visualisierungen. Sicherheitsaspekte wie Zugriffskontrollen, Datenverschlüsselung und die Einhaltung von Compliance-Vorgaben sollten dabei ebenfalls berücksichtigt werden.

\begin{itemize}
\item \textbf{Langfuse}~\cite{langfuse}: Ein spezialisiertes Monitoring-Tool für LLM-Systeme, das Einblick in Inferenzzeiten, Tokenverbrauch und Modellantworten bietet und so die Analyse von Modell-Performance und Fehlerraten erleichtert.
\item \textbf{Prometheus}~\cite{prometheus} und \textbf{Grafana}~\cite{grafana}: Ein leistungsstarkes Duo zur Erfassung, Speicherung und Visualisierung von Metriken; Prometheus sammelt und verarbeitet Echtzeitdaten, während Grafana anpassbare Dashboards für die übersichtliche Darstellung bietet.
\item \textbf{ELK Stack} (Elasticsearch, Logstash, Kibana)~\cite{elk}: Eine umfassende Lösung für Log-Management und -Analyse; Logstash verarbeitet und speichert Logs in Elasticsearch, die in Kibana visualisiert werden können.
\item \textbf{Loki}~\cite{loki}: Ein skalierbares, kosteneffizientes Logging-System, das sich nahtlos in Grafana integriert und die Aggregation sowie Suche von Logs ermöglicht, ohne hohe Speicheranforderungen zu stellen.
\item \textbf{Jaeger}~\cite{jaeger}: Ein Tool für verteiltes Tracing, das die Nachverfolgung von Anfragen über verschiedene Microservices hinweg erleichtert und somit bei der Identifizierung von Latenzproblemen und Engpässen hilft.
\item \textbf{Sentry}~\cite{sentry}: Eine Plattform zur Fehlerverfolgung und Benachrichtigung, die Echtzeitwarnungen bei Code-Fehlern liefert und die Analyse von Abstürzen, Performance-Problemen und Bugs unterstützt.
\end{itemize}

\subsubsection{Deployment und Orchestrierung}

Das Deployment und die Orchestrierung von Anwendungen umfassen die Bereitstellung, Verwaltung und Skalierung von Software in verschiedenen Umgebungen. Ziel ist es, Anwendungen effizient, zuverlässig und reproduzierbar bereitzustellen. Während das Deployment den Prozess der Auslieferung von Software beschreibt, sorgt die Orchestrierung für die Koordination mehrerer Dienste, deren Skalierung, Lastverteilung und Fehlertoleranz. Dies ist besonders in komplexen Systemarchitekturen und bei Microservices entscheidend, um einen stabilen und performanten Betrieb sicherzustellen.

Wichtige Kriterien für Deployment- und Orchestrierungslösungen sind eine einfache Skalierbarkeit, um Anwendungen flexibel an wechselnde Lasten anzupassen, sowie eine hohe Verfügbarkeit durch automatische Wiederherstellungsmechanismen und Lastverteilung. Ebenso wichtig sind Automatisierungsmöglichkeiten zur Unterstützung von Continuous Integration/Continuous Deployment (CI/CD), um Entwicklungs- und Bereitstellungsprozesse zu beschleunigen. Eine benutzerfreundliche Verwaltung der Infrastruktur sowie die Integration von Sicherheitsmechanismen wie Zugriffskontrollen und Secret-Management sind ebenfalls entscheidend.

Zur Umsetzung stehen verschiedene Open-Source-Tools zur Verfügung:

\begin{itemize}
\item \textbf{Kubernetes}~\cite{kubernetes}: Eine führende Plattform zur Orchestrierung von Containern, die automatische Skalierung, Selbstheilung und Service-Discovery ermöglicht.
\item \textbf{Docker}~\cite{docker}: Ein Tool zur Containerisierung von Anwendungen, das Portabilität über verschiedene Umgebungen hinweg sicherstellt.
\item \textbf{Helm}~\cite{helm}: Ein Paketmanager für Kubernetes, der die Verwaltung komplexer Deployments durch wiederverwendbare Konfigurationspakete erleichtert.
\item \textbf{Argo CD}~\cite{argo}: Ein GitOps-Tool für kontinuierliche Bereitstellung, das Deployments automatisch mit dem Quellcode synchronisiert.
\item \textbf{Kubeflow}~\cite{kubeflow}: Eine Plattform zur Orchestrierung von Machine-Learning-Workflows, die Modelltraining und Inferenzprozesse unterstützt.
\end{itemize}

\subsubsection{Testing und Evaluation}

Zur Sicherstellung der Qualität und Zuverlässigkeit des Systems ist eine umfassende Teststrategie unerlässlich. Diese soll sicherstellen, dass alle Komponenten fehlerfrei funktionieren, erwartete Ergebnisse liefern und unter verschiedenen Bedingungen stabil bleiben. Wichtige Kriterien für das Testing sind die Abdeckung verschiedener Testebenen, einschließlich Unit-, Integrations- und End-to-End-Tests, um sowohl einzelne Funktionen als auch das Zusammenspiel der Systemkomponenten zu überprüfen. Ebenso wichtig ist die Automatisierung der Tests, um Entwicklungszyklen zu beschleunigen und die Wiederholbarkeit zu gewährleisten. Für datengetriebene Systeme spielt außerdem die Validierung der Datenqualität und die Modellbewertung eine zentrale Rolle.  

Dabei können die folgenden Open-Source-Projekte eingesetzt werden:

\begin{itemize}
\item \textbf{DeepEval}~\cite{deepeval}: Ein Tool zur Bewertung von LLM-Ausgaben und zur automatisierten Modellbewertung.
\item \textbf{pytest}~\cite{pytest}: Ein weit verbreitetes Framework für Unit- und Integrationstests in Python mit einfacher Syntax und umfangreichen Erweiterungsmöglichkeiten.
\item \textbf{Robot Framework}~\cite{robot}: Ein plattformübergreifendes Testautomatisierungs-Framework, das End-to-End- und Akzeptanztests unterstützt.
\item \textbf{Great Expectations}~\cite{greatexpectations}: Ein Tool zur Validierung der Datenqualität, das Datenpipelines überwacht und sicherstellt, dass Daten den definierten Erwartungen entsprechen.
\item \textbf{MLflow}~\cite{mlflow}: Eine Plattform zur Verfolgung von Machine-Learning-Experimenten, Modellversionierung und Bewertung von Modellen über verschiedene Trainingsläufe hinweg.
\end{itemize}

\subsubsection{Zusätzliche Komponenten}

Neben den Kernkomponenten können weitere Open-Source-Tools integriert werden, um die Funktionalität, Robustheit und Skalierbarkeit des Systems zu erweitern. Diese ergänzenden Komponenten ermöglichen eine bessere Verwaltung von APIs, Modellen und Workflows sowie eine effiziente Sicherung und Wiederherstellung von Daten. Wichtige Kriterien bei der Auswahl solcher Tools sind die einfache Integration in bestehende Architekturen, Skalierbarkeit, Sicherheitsfunktionen und die Fähigkeit, Betriebsprozesse zu automatisieren.

Zur Auswahl stehen u.a. die folgenden Lösungen:

\begin{itemize}
\item \textbf{API-Gateways}: \textbf{Kong}~\cite{kong} und \textbf{Tyk}~\cite{tyk} ermöglichen die sichere Verwaltung von API-Endpunkten, inklusive Authentifizierung, Rate-Limiting und Lastverteilung.
\item \textbf{Modellverwaltung}: \textbf{DVC}~\cite{dvc} und \textbf{MLflow}~\cite{mlflow} bieten Funktionen zur Versionskontrolle von Modellen, zur Nachverfolgung von Experimenten und zur Reproduzierbarkeit von Trainingsprozessen.
\item \textbf{Workflow-Management}: \textbf{Apache Airflow}~\cite{airflow} unterstützt die Planung, Überwachung und Automatisierung von Datenpipelines und wiederkehrenden Prozessen.
\item \textbf{Backup und Wiederherstellung}: \textbf{Velero}~\cite{velero} bietet eine zuverlässige Lösung zur Sicherung, Wiederherstellung und Migration von Kubernetes-Cluster-Ressourcen und -Daten.
\end{itemize}

Die Integration dieser Komponenten verbessert die Betriebsstabilität, erhöht die Effizienz und stellt sicher, dass das System auch unter wachsenden Anforderungen skalierbar und wartbar bleibt.

\section{Diskussion}
\label{sec:discussion}

In den vorangegangenen Kapiteln haben wir die Motivation, die Herausforderungen und eine detaillierte Referenzarchitektur für die Entwicklung von geschlossenen LLM-Systemen vorgestellt. In diesem Kapitel sollen die vorgestellten Konzepte kritisch reflektiert und im Kontext aktueller Entwicklungen diskutiert werden. Dabei werden sowohl die Vorteile als auch die potenziellen Einschränkungen der vorgeschlagenen Architektur beleuchtet.

\subsection{Bewertung der Architektur}

Die vorgeschlagene Referenzarchitektur bietet eine Vielzahl an Vorteilen, bringt jedoch auch Herausforderungen mit sich, die bei der Implementierung berücksichtigt werden müssen.

\paragraph{Datenschutz und Sicherheit}

Durch die Implementierung als geschlossenes System wird der Datenschutz erheblich gestärkt. Daten verbleiben innerhalb der kontrollierten Infrastruktur, wodurch das Risiko von Datenschutzverletzungen und unautorisiertem Zugriff minimiert wird. Die Nutzung von Open-Source-Sicherheitskomponenten wie Llama Guard ~\cite{llamaguard} und Guardrails~\cite{guardrails} ermöglicht es, Sicherheitsmechanismen transparent zu gestalten und an spezifische Anforderungen anzupassen. Gleichzeitig erfordert die eigenverantwortliche Verwaltung eines geschlossenen Systems umfassende Sicherheitskonzepte, regelmäßige Audits und Aktualisierungen, um Sicherheitsrisiken zu minimieren.

\paragraph{Flexibilität und Open-Source-Vorteile}

Die modulare Architektur erlaubt es, Komponenten auszutauschen oder zu erweitern. Dies bietet Flexibilität bei der Integration neuer Technologien und bei der Anpassung an neue oder geänderte Anforderungen. Die Verwendung von Open-Source-Tools fördert zudem die Transparenz und Anpassbarkeit, da der Quellcode verfügbar ist und Modifikationen ermöglicht. Allerdings kann der Aufbau eines stabilen Open-Source-Stacks mit hohem Initialaufwand verbunden sein, insbesondere in Bezug auf die Kompatibilität zwischen verschiedenen Tools.

\paragraph{Technische Komplexität und Wartung}

Die Integration und Wartung einer offenen Architektur erfordert technisches Fachwissen und kontinuierliche Betreuung. Unterschiedliche Open-Source-Tools und neuere Versionen können inkompatible Schnittstellen oder Abhängigkeiten aufweisen, was zusätzlichen Entwicklungsaufwand mit sich bringt. Der Wartungsaufwand ist nicht zu unterschätzen, da regelmäßige Updates erforderlich sind, um Sicherheitslücken zu schließen und neue Funktionen zu integrieren. Ohne eine klare Wartungsstrategie kann die langfristige Nutzung erschwert werden.

\subsection{Vergleich mit alternativen Ansätzen}

Die vorgeschlagene Architektur unterscheidet sich von anderen möglichen Implementierungen, insbesondere durch ihren Open-Source-Fokus und die dezentrale Datenverarbeitung.

\paragraph{Proprietäre vs. Open-Source-Modelle}

Proprietäre LLM-Lösungen bieten oft eine schnellere und einfachere Integration in eigene Prozesse sowie den kommerziellen Support durch den jeweiligen Anbieter. Allerdings ist dies mit entsprechenden, teils hohen Lizenzkosten verbunden und brigt zudem das Risiko eines Vendor Lock-ins. Open-Source-Modelle bieten Unabhängigkeit und Transparenz, erfordern jedoch mehr Eigenverantwortung hinsichtlich Wartung, Sicherheit und Weiterentwicklung.

\paragraph{Cloud- vs. On-Premise-Strategien}

Cloud-basierte LLM-Dienste ermöglichen eine schnelle Skalierung und einfache Bereitstellung, bringen jedoch  Datenschutzrisiken und eine hohe Abhängigkeit von gewählten Cloud-Anbietern mit sich. Durch den Einsatz von Sicherheitsmechanismen wie wie Ende-zu-Ende-Verschlüsselung, strenge Zugriffskontrollen mit minimalen Zugriffsberechtigungen kann das Datenschutzrisiko reduziert werden. Geschlossene On-Premise-Lösungen hingegen bieten eine vollständige Kontrolle über Daten und Modelle, erfordern jedoch erhebliche Investitionen in IT-Infrastruktur. Hybride Ansätze, bei denen weniger kritische Daten in der Cloud verarbeitet werden, während sensible Informationen On-Premise bleiben, könnten eine praktikable Alternative sein. Die Kombination aus hybriden Ansätzen und modernen Verschlüsselungsmethoden kann eine ausgewogene Lösung für Datenschutz und Skalierbarkeit bieten.

\paragraph{Praxisbeispiele für geschlossene LLM-Systeme}

In regulierten Branchen wie dem Gesundheitswesen oder der Finanzbranche setzen Unternehmen zunehmend auf geschlossene KI-Systeme, um strenge Datenschutzrichtlinien einzuhalten. Beispielsweise gibt es verschiedene Ansätze und Bestrebungen, geschlossene LLMs zur Unterstützung von medizinischen Diagnosen sowie der Erstellung von Berichten oder für Compliance-Überprüfungen in der Finanzbranche zu nutzen. Diese Anwendungen zeigen, dass die Umsetzung eines geschlossenen Systems realistisch ist, jedoch hohe Anforderungen an Sicherheit, Skalierbarkeit und Effizienz stellt.

\subsection{Zukunftsperspektiven und Empfehlungen}

Die kontinuierliche Weiterentwicklung von LLMs und Open-Source-Technologien wird maßgeblich beeinflussen, wie geschlossene Systeme in Zukunft gestaltet werden können.

\paragraph{Verbesserte Open-Source-Modelle}

Die Entwicklungen im Bereich der Open-Source-Modelle verlaufen derzeit äußerst dynamisch, sodass in naher Zukunft mit weiteren signifikanten Fortschritten zu rechnen ist. Mit der Veröffentlichung neuer Modelle wie LLaMa3~\cite{llama3} oder Mistral~\cite{mistral} steigt die Qualität verfügbarer Open-Source-Modelle, wodurch geschlossene Systeme effizienter betrieben werden können, ohne auf proprietäre Lösungen angewiesen zu sein.

\paragraph{Automatisierung und MLOps}

Die Integration von MLOps-Praktiken kann den Entwicklungs- und Wartungsprozess optimieren. Automatisierte Tests, Deployment-Workflows und Monitoring-Tools helfen, den Betrieb eines geschlossenen LLM-Systems langfristig stabil zu halten und Updates effizient umzusetzen.

\paragraph{Praktische Herausforderungen bei der Umsetzung}

Eine der größten Herausforderungen bleiben die technische Umsetzung und besonders die Wartung. Unternehmen sollten daher frühzeitig eine Strategie für den Aufbau, die Wartung und die Weiterentwicklung ihres geschlossenen Systems definieren. Hierzu gehören regelmäßige Sicherheitsüberprüfungen, klare Zuständigkeiten innerhalb des Teams und die Auswahl langfristig unterstützter Open-Source-Technologien.

\paragraph{Langfristige Strategien für Wartung und Anpassung}

Eine nachhaltige Architektur sollte so gestaltet sein, dass sie leicht aktualisierbar ist und mit neuen Entwicklungen Schritt halten kann. Unternehmen sollten eine kontinuierliche Evaluierung der eingesetzten Modelle und Tools vornehmen, um sicherzustellen, dass ihr System langfristig effizient und sicher bleibt. 

\paragraph{Best Practices für eine erfolgreiche Umsetzung}

Basierend auf den Diskussionsergebnissen ergeben sich folgende Empfehlungen für eine erfolgreiche Implementierung in Unternehmen, um geschlossene LLM-Systeme effizient und zukunftssicher gestalten können:

\begin{itemize}
    \item \textbf{Gründliche Planung}: Eine umfassende Anforderungsanalyse und Pilotprojekte helfen, potenzielle Herausforderungen frühzeitig zu identifizieren. Pilotprojekte ermöglichen agile, iterative Tests, bei denen Ansätze schnell evaluiert und entweder weiterentwickelt oder verworfen werden können.
    \item \textbf{Investition in Personal und Schulung}: Technisches Know-how ist entscheidend für eine erfolgreiche Implementierung und Wartung.
    \item \textbf{Kombination aus Automatisierung und regelmäßigen Audits}: Automatisierte Prozesse zur Wartung und Sicherheitsüberprüfung können langfristig den Betrieb erleichtern.
    \item \textbf{Regelmäßige Evaluierung neuer Open-Source-Modelle}: Die KI-Landschaft entwickelt sich rasant, daher sollte das System flexibel genug sein, um neue Modelle und Technologien zu integrieren.
\end{itemize}

\section{Zusammenfassung}
\label{sec:conclusion}

Die steigende Verbreitung von Large Language Models (LLMs) bringt großes Potential aber auch erhebliche Herausforderungen im Bereich Datenschutz, Sicherheit und regulatorischer Compliance mit sich. Während offene, cloudbasierte LLM-Dienste oft leistungsfähige Lösungen bieten, stehen sie im Widerspruch zu den strengen Anforderungen, die Unternehmen in regulierten Branchen erfüllen müssen. Die Entwicklung geschlossener LLM-Systeme stellt eine vielversprechende Alternative dar, erfordert jedoch eine durchdachte Architektur, die sowohl technische als auch organisatorische Aspekte berücksichtigt.

In dieser Arbeit präsentieren wir eine Open-Source-Referenzarchitektur für geschlossene LLM-Systeme, die Datenschutz und Sicherheit mit technischer Skalierbarkeit und Flexibilität verbindet. Unser Ansatz kombiniert moderne Technologien für effiziente Datenverwaltung, Inferenzoptimierung und Sicherheitsmechanismen. Die Architektur nutzt Open-Source-Modelle, um Vendor Lock-in zu vermeiden und erlaubt durch modulare Komponenten eine anwendungsfallspezifische Anpassung. Besondere Schwerpunkte liegen auf der sicheren Verarbeitung sensibler Daten und der zukunftsfähigen Systemintegration.

Die Analyse zeigt, dass geschlossene LLM-Systeme in der Praxis realisierbar sind, jedoch mit Herausforderungen in Bezug auf Rechenressourcen, Skalierbarkeit und Wartung verbunden sind. Unsere vorgeschlagene Architektur bietet Lösungsansätze, um diese Herausforderungen zu bewältigen. Zudem können alternative Betriebsmodelle, darunter hybride Architekturen, eine Balance zwischen Datenschutz und Skalierbarkeit ermöglichen.

Zukünftige Forschungsarbeiten sollten sich mit der empirischen Evaluierung der vorgeschlagenen Architektur, der weiteren Untersuchung ressourcenschonender Modelle sowie der Integration neuer regulatorischer Anforderungen befassen. Die vorliegende Arbeit liefert eine Grundlage für Organisationen, die eine datenschutzkonforme und nachhaltige Nutzung von LLMs anstreben.

\renewcommand\refname{Referenzen}
\bibliographystyle{plain}
\bibliography{references} 

\end{document}